\documentclass[a4paper,12pt,oneside,titlepage]{article}
\usepackage[cp1251]{inputenc}
\usepackage[T2A]{fontenc}
\usepackage{indentfirst}
\usepackage{amsmath}
\usepackage{amsxtra}
\usepackage{amssymb}
\usepackage{longtable}
\usepackage{graphicx}
\usepackage{cite}
\usepackage{psfrag}

\newcommand{\eps}{\varepsilon}
\newcommand{\kap}{\varkappa}

\usepackage[small]{caption2}
\large  

\renewcommand{\thesection}{\arabic{section}}

\numberwithin{equation}{section} \numberwithin{figure}{section}
\numberwithin{table}{section}

\begin{document}

\begin{center}
{\huge Plasma opacity calculations using the Starrett and Saumon average-atom model with ion correlations}

\vskip 1cm
A.A. Ovechkin$^{a}$\footnote{ovechkin.an@mail.ru}, P.A. Loboda$^{a,b}$, A.L. Falkov$^{a,b}$

\vskip 0.5cm
{\small{$^{a}$Russian Federal Nuclear Center --- All-Russian Research Institute of Technical Physics (RFNC-VNIITF), 13, Vasilyeva st., Snezhinsk, Chelyabinsk region, 456770, Russia

$^{b}$National Research Nuclear University --- Moscow Engineering Physics Institute (MEPhI), 31, Kashirskoe sh., Moscow, 115409, Russia}}

\vskip 1cm
\textbf{Abstract}
\end{center}

We present the opacities of iron, aluminum, and bromine plasmas calculated using the Starrett and Saumon average-atom model allowing for ion correlations. We show that the use of earlier average-atom ion-correlation model of Rozsnyai, as has recently been done in the solar opacity calculations, overestimates the effect of ion correlations on plasma opacities. The reason for this overestimation is discussed.


\section{Introduction}

The knowledge of radiative properties of hot plasmas plays a key role in characterizing heat transfer and dynamic processes specific to many problems of high-energy-density physics and astrophysics. Over the past decade, much effort has been devoted to the problem of solar opacity. The problem is that the use of the revised abundances of heavy elements in the solar mixture \cite{Asplund2009} brought the predictions of standard solar models utilizing up-to-date theoretical opacity data into disagreement with experimental data from helioseismic observations and neutrino measurements. This disagreement may be eliminated if the Rosseland mean opacities \cite{ZRe} of the solar mixture would in fact be greater by a value ranging from approximately 5 \% in the center of the Sun to $\simeq 20$ \% around the inner boundary of its convection zone \cite{Christensen_Dalsgaard2009,Villante2010,Villante2015}. Such a modest increase of the solar-mixture Rosseland means would imply much more pronounced enhancement of the abundant heavy-element (C, N, O, Ne, \ldots, Fe, Ni) Rosseland means. This conjecture is supported by the recent measurements of iron transmission spectra performed at the Sandia Z machine \cite{Bailey2015} at a temperature $T\simeq 180$ eV, being specific to the convection and radiative zone boundary, and an electron density approximately 2.5 times lower than that one anticipated at this boundary.

The measurements showed that monochromatic opacities in the $L$-shell photoionization spectral range are twice as high as the relevant predictions by the known up-to-date theoretical models. It is important to note that detailed calculations of solar opacities using those models, e.g. ATOMIC \cite{Colgan2013,Colgan2015,Colgan2016}, OPAS \cite{Blancard2012,Mondet2015,LePennec2015}, and SCO-RCG \cite{Porcherot2011,Pain2015_2,Pain2017} as well as the earlier OP (Opacity Project) \cite{Seaton2004,Seaton2005,Badnell2005,Seaton2007} and OPAL \cite{Iglesias1996} calculations, yield similar results, being however unable to explain the data of helioseismic observations and neutrino measurements. This contradiction along with the unexpected results of the Z-machine experiment \cite{Bailey2015} have spurred the efforts to reveal further potential improvements of theoretical opacity data \cite{Krief2016_2,Nahar2016,Pain2017}. One such improvement was associated with a more consistent description of plasma density effects. Specifically, this could be done by generating all the necessary atomic data (intra-atomic interaction and transition matrix elements) with an average-atom model allowing for ion correlations.

Earlier opacity calculations with ion correlation effects included \cite{Rozsnyai1991,Rozsnyai1992,Krief2018}
involved average-atom ion-correlation model of Rozsnyai \cite{Rozsnyai1991,Rozsnyai2014}. Among those, the most recent calculations \cite{Krief2018} performed with the STAR code \cite{Krief2015,Krief2015_2,Krief2016,Krief2016_2} implementing the statistical superconfiguration STA model \cite{STA1989,Bar_Shalom1995,Bar_Shalom1996} have revealed an enhancement of the solar-mixture Rosseland means due to ion correlations ranging from 1.5 -- 2 \% in the center of the Sun up to 10 \% at its convection zone boundary. This is however nearly twice as low as one would need to drive the solar-model predictions into agreement with the data of helioseismic observations and neutrino measurements. Though the results of Ref. \cite{Krief2018} do not provide a reasonable solution of the solar-opacity problem, the confirmation of those would evidently mean a tangible progress in this direction.

In section {\ref{Rozsnyai_section}} of the present paper we examine the effects of ion correlations on calculated plasma opacities found in earlier work \cite{Rozsnyai1991,Rozsnyai1992,Krief2018}. In section {\ref{Starrett_section}} we provide opacity calculations done with an ion-correlation average-atom model recently formulated by Starrett and Saumon \cite{Starrett2013,Starrett2014} and make comparisons with the results of earlier work for some representative cases of Refs. \cite{Rozsnyai1991,Krief2018}. Final section briefly formulates our conclusions.

\section{The effect of ion correlations on plasma opacities}\label{Rozsnyai_section}

In the majority of up-to-date versions of average-atom model \cite{Blenski2007_VAAQP_2,Blenski2013_VAAQP} considering the average ion embedded in an infinite jellium of screening plasma electrons at a neutralizing ion background, like Liberman's model \cite{Liberman1979}, the neutral Wigner-Seitz-sphere (NWS) model \cite{Piron2011_VAAQP} or the VAAQP model \cite{Blenski2007_VAAQP,Piron2011_VAAQP}, plasma ions of average ion density $n_{i}$ cannot penetrate into any specific atomic cell (Wigner-Seitz sphere) of radius
\begin{equation}\label{cell_radius}
r_{0}=\left(\dfrac{3}{4\pi\,n_{i}}\right)^{1/3}, \quad n_{i}=N_{A}\,\rho/A,
\end{equation}
where $\rho$ is the material density, $N_{A}$ is the Avogadro constant, and $A$ is the atomic weight (in g/mole). In those models, the spatial distribution of plasma ions external to the central ion of the atomic cell is also supposed to be uniform. Therefore, the relevant ion-ion pair correlation function is just
\begin{equation}\label{gii_0}
g(r)=\theta(r-r_{0}),
\end{equation}
where $\theta(x)$ is the Heaviside function.

Ion-correlation average-atom models allow for the non-uniformity of spatial distribution of all external charges (both free electrons and ions) thus enabling one to replace the step function (\ref{gii_0}) by a more realistic distribution. The use of the latter one therefore provides a capability to consistently describe the effect of ion correlations on the electron energy levels and wave functions. This effect manifests itself in three principal ways. First, it results in the alteration of the electron-configuration probabilities as, for instance, may be seen in earlier calculations of Ref. \cite{Rozsnyai1991} done for aluminum plasma at a temperature $T=100$ eV and material density $\rho=0.0045$ g/cm$^{3}$. Specifically, an average-atom model allowing for ion correlations (IC) yielded the value of mean ion charge (the difference of nuclear charge and mean number of bound electrons) $Z_{0}^{(IC)}=10.63$ \cite{Rozsnyai1991} slightly different from that one obtained with ion-sphere-restricted (IS) average-atom model disregarding ion correlations --- $Z_{0}^{(IS)}=10.76$ \cite{Rozsnyai1991}. Since the opacity in this case is dominated by the bound-free and bound-bound absorption, it is more informative here to compare mean numbers of bound electrons, $\langle Q\rangle=Z-Z_{0}$, taking the values $\langle Q\rangle^{(IC)}=2.37$ and $\langle Q\rangle^{(IS)}=2.24$ with the use of the IC and IS models, respectively.

At plasma conditions considered, the IC and IS models both showed the aluminum $K$-shell to be essentially closed. In this case, the $K$-shell transition energies $\omega\gtrsim1.5$ keV are far beyond the spectral region around the energy $\omega\approx 3.8T=380$ eV in which the Rosseland weighting function maximizes. As a result, the Rosseland mean opacity is mostly governed by the number of bound electrons occupying all other shells except the $K$-shell, $\langle \tilde{Q}\rangle \approx \langle Q\rangle-2$, so that the relevant values obtained with the IS and IC models differ by a factor of 1.5: $\langle \tilde{Q}\rangle $ changes from $\langle\tilde{Q}\rangle^{(IS)}=0.24$ to $\langle\tilde{Q}\rangle^{(IC)}=0.37$. Consequently, the IC model provides higher probabilities of the configurations involving the $L$-, $M$-, $\ldots$ shell electrons than the IS model does. This, in turn, yields as much larger value of the Rosseland mean opacity $\kap_{R}$ changing from $\kap_{R}^{(IS)}=84.3$ cm$^{2}$/g to $\kap_{R}^{(IC)}=131$ cm$^{2}$/g \cite{Rozsnyai1991}. So, a seemingly minor decrease of mean ionization due to ion correlations in Ref. \cite{Rozsnyai1991} resulted in the strong enhancement of the Rosseland mean opacity.

The second principal manifestation of the ion-correlation effect on the electron properties is the alteration of the bound-bound and bound-free oscillator strengths. This alteration has recently been examined in Ref. \cite{Krief2018} by analyzing monochromatic opacity of iron calculated for the conditions of the Z-machine opacity measurements \cite{Bailey2015}: $T=182$ eV, $\rho=0.17$ g/cm$^{3}$ (mean free electron density $n_{e}=3.1\cdot10^{22}$ cm$^{-3}$). Specifically, the inclusion of ion correlations caused a pronounced enhancement of the $M$-shell photoionization cross-section that yielded the 20 \% increase of the Rosseland mean opacity (from 605.03 cm$^{2}$/g to 725.49 cm$^{2}$/g) as it was sensitive to the $M$-shell bound-free photoabsorption in the case considered. At the same time, monochromatic opacity remained almost intact in the spectral range of the $L$-shell photoabsorption in which the opacity measurements in the Z-machine experiment were done \cite{Bailey2015}.

Finally, it is shown below that the effect of ion correlations on plasma opacities may also manifest itself through a non-negligible shift of spectral-line arrays and photoionization thresholds to higher photon energies.

\section{Opacity calculations using the Starrett and Saumon model}\label{Starrett_section}

\begin{figure}
 \begin{center}
 \psfrag{xlab}{$\omega$, eV}
 \psfrag{ylab}{$\kap(\omega)$, cm$^{2}$/g}
 \psfrag{b}{\textit{a}}
 \psfrag{a}{\textit{b}}
 \psfrag{c}{\textit{c}}
 \includegraphics[scale=0.42,angle=270]{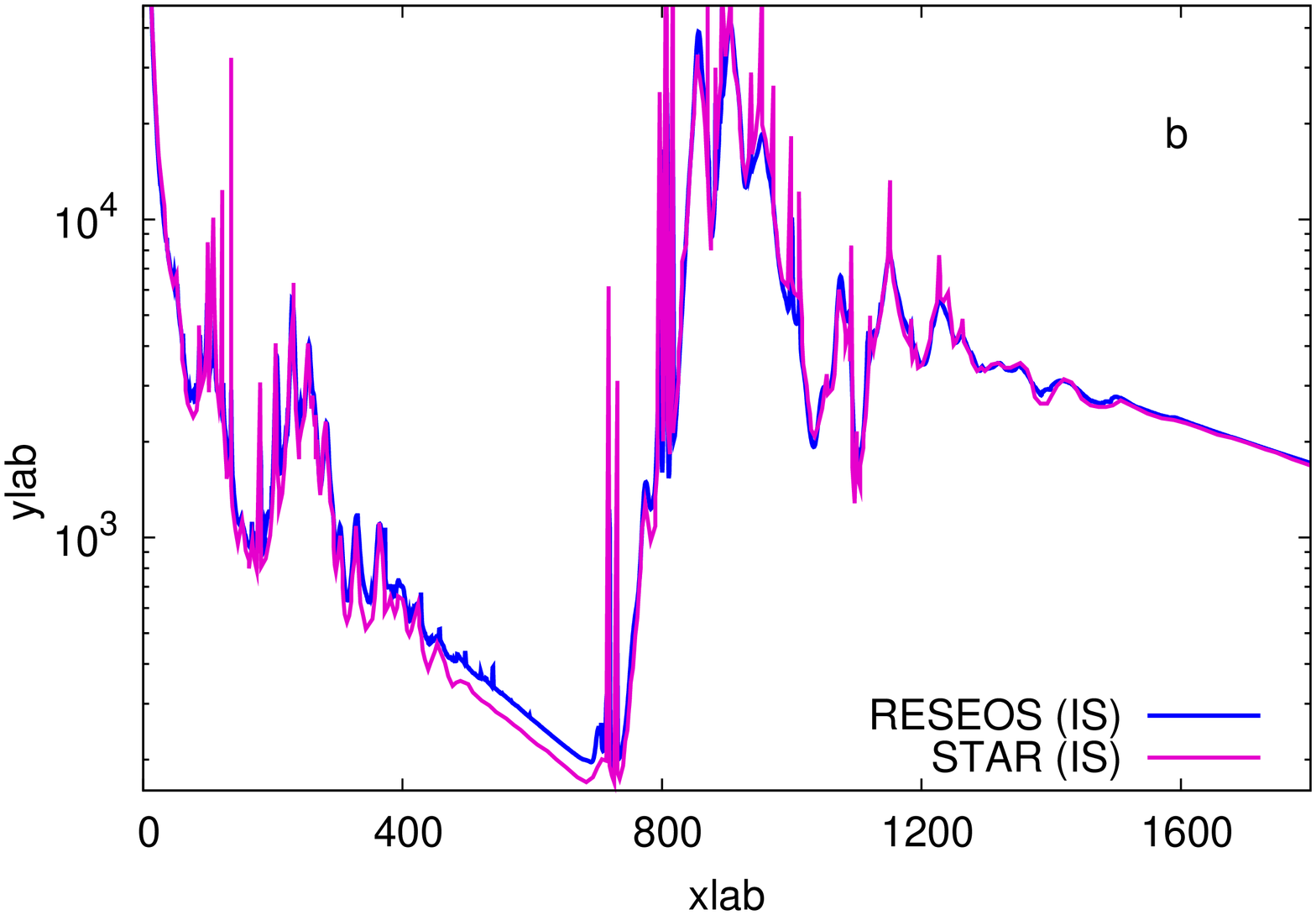}
 \includegraphics[scale=0.42,angle=270]{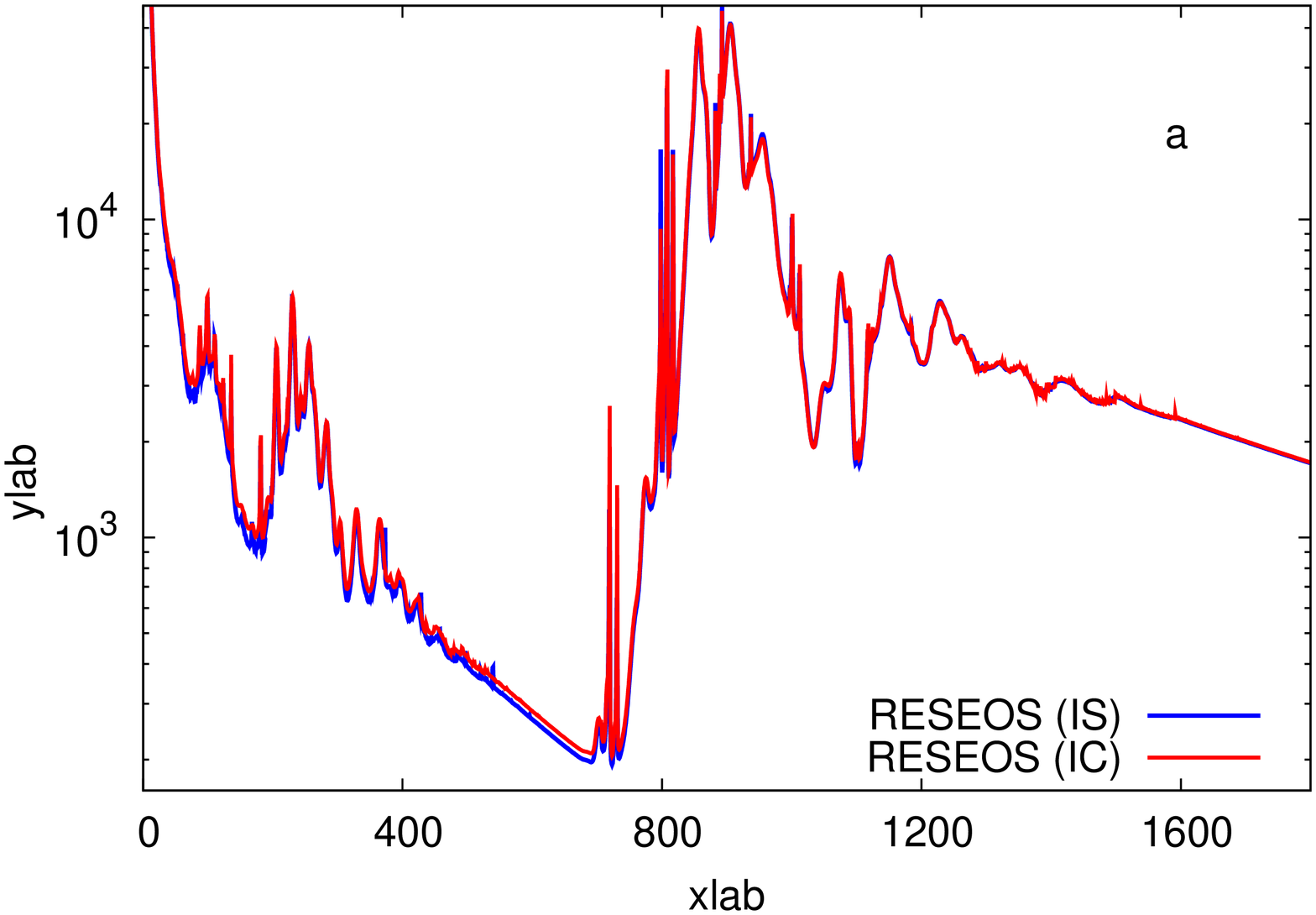}
 \includegraphics[scale=0.42,angle=270]{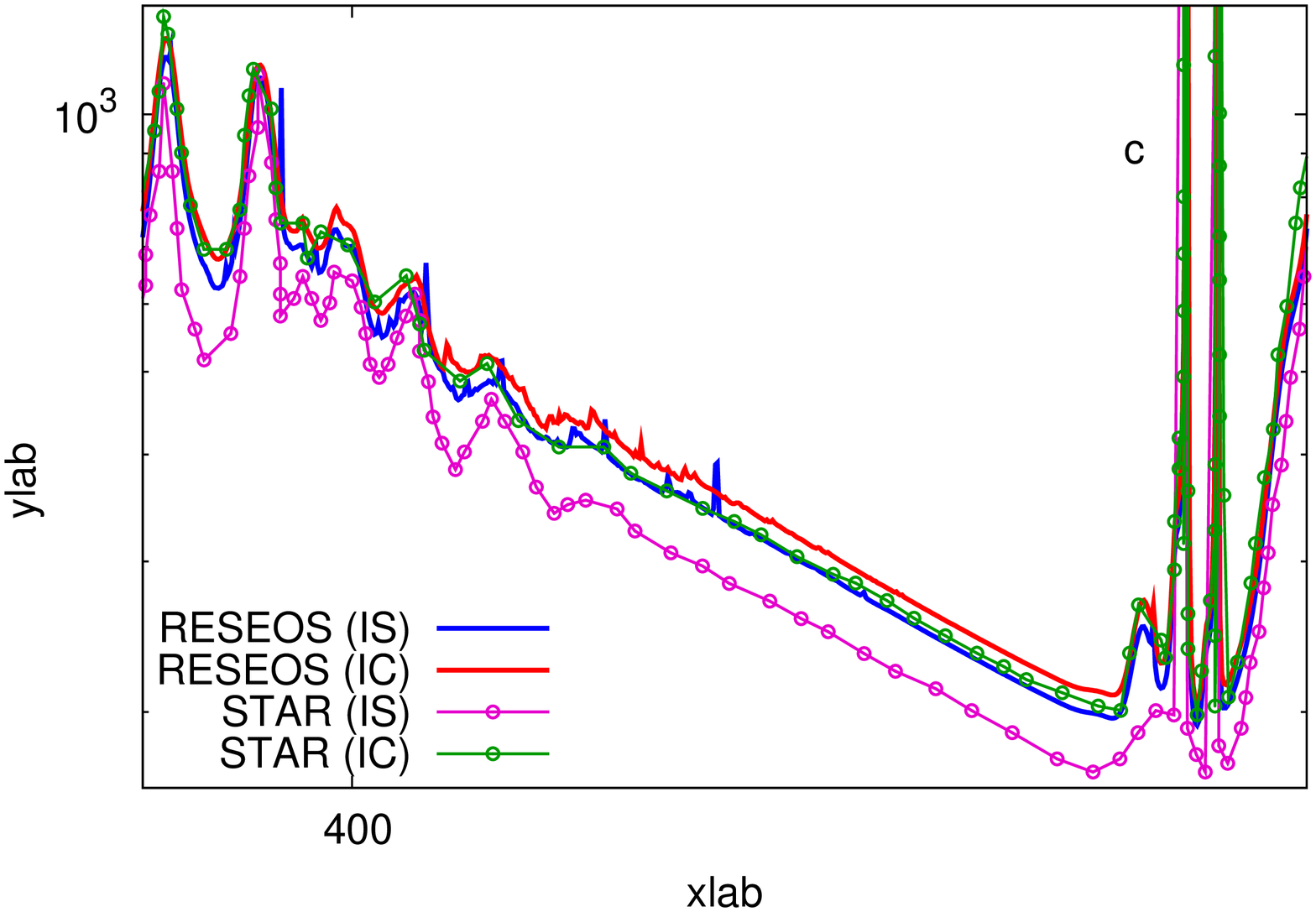}
 \caption{The monochromatic opacity of iron at a temperature $T=182$ eV and material density $\rho=0.17$ g/cm$^{3}$ ($n_{e}=3.1\cdot10^{22}$ cm$^{-3}$) calculated using the RESEOS code with (red curve, $\kap_{R}=743.06$ cm$^{2}$/g) and with no (blue curve, $\kap_{R}=704.12$ cm$^{2}$/g) regard for ion correlations as compared to the corresponding STAR-code calculations \cite{Krief2018} with (green curve, $\kap_{R}=725.49$ cm$^{2}$/g) and with no (magenta curve, $\kap_{R}=605.03$ cm$^{2}$/g) regard for ion correlations.}
 \label{Spectr_Fe_STAR_fig}
 \end{center}
\end{figure}

\begin{figure}
 \begin{center}
 \psfrag{xlab}{$\eps$, eV}
 \psfrag{ylab}{$f_{\alpha,\,\eps lj}$}
 \includegraphics[scale=0.45,angle=270]{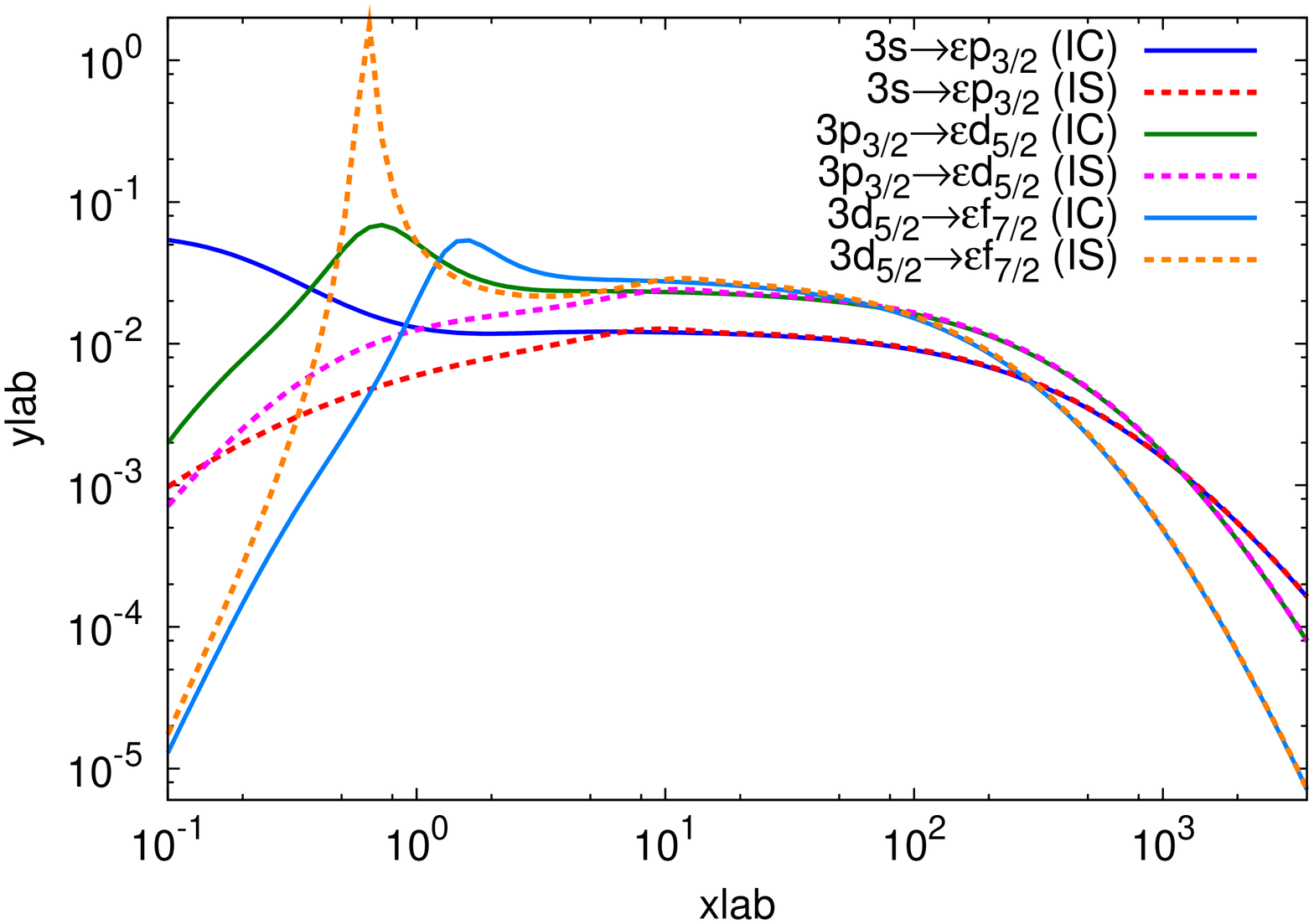}
 \caption{Single-electron oscillator strengths for some bound-free $M$-shell dipole transitions in iron at $T=182$ eV, $\rho=0.17$ g/cm$^{3}$ ($n_{e}=3.1\cdot10^{22}$ cm$^{-3}$) as calculated by RESEOS in relativistic velocity form with (IC model, solid curves) and with no (IS model, dashed curves) regard for ion correlations.}
 \label{oscil_Fe_fig}
 \end{center}
\end{figure}

To reveal the effect of ion correlations on plasma opacities we performed the calculations using an ion-correlation average-atom model recently formulated by Starrett and Saumon \cite{Starrett2013,Starrett2014}. The model was implemented in the RESEOS code \cite{RESEOS2014,RESEOS2016} that employs a generalized version of the STA model \cite{RESEOS2014} providing
a substantial acceleration of the bound-bound and bound-free absorption calculations with practically the same accuracy as that one obtained with the original superconfiguration approach \cite{STA1989,Bar_Shalom1995,Bar_Shalom1996}. In present paper, all the ion-correlation calculations are done in the first-order approximation implying a two-stage solution of the electronic-structure equations with the NWS model at the first stage and the IC model \cite{Starrett2013,Starrett2014} at the second one. Correlation functions for the IC-model calculations are found only once by using the NWS-model electronic data. In the framework of the Starrett and Saumon model, this approach alone yields the results very similar to those ones obtained with fully self-consistent calculations of electronic and ionic structures \cite{Starrett2013,Starrett2014}.

Unlike the Liberman model \cite{Liberman1979} originally implemented in RESEOS \cite{Sinko2013,RESEOS2014}, the NWS model allows for non-uniformity of the electron density at $r>r_{0}$ while retaining the pair correlation function in the form of the step function (\ref{gii_0}). However, RESEOS calculations show that in the low-density cases considered below the NWS and Liberman models provide nearly coincident results since the electron density in the NWS model at $r>r_{0}$ becomes almost uniform at high temperatures and low material densities (see, e.g., Ref. \cite{RESEOS2016} and Appendix A). In these conditions, both of the models actually become equivalent to the IS model. Therefore, the RESEOS NWS calculations are hereafter denoted as the IS ones.

First, we calculated the iron opacity at the same conditions as in Ref. \cite{Krief2018} ($T=182$ eV, $\rho=0.17$ g/cm$^{3}$). One can see from Fig. {\ref{Spectr_Fe_STAR_fig}}\textit{a} that the IS calculations done with the RESEOS and STAR codes yield very close results almost everywhere except for the spectral range of $400 - 750$ eV. In this range, the $M$-shell photoionization cross-section calculated with RESEOS appears to be larger than that one given by the STAR code, thus being responsible for the 16 \% difference of the relevant Rosseland means. This difference is however not of primary importance here as we are mostly interested in relative alterations of the Rosseland means due to ion correlations.

The use of the Starrett and Saumon model offered slightly smaller mean ion charge ($Z_{0}^{(IC)}=16.66$) than that one obtained with the NWS model ($Z_{0}^{(IS)}=16.88$). The relevant minor enhancement of the average $M$-shell occupation number of 2.6 \% is evidently insufficient to provide the 20 \% increase of the Rosseland mean opacity due to higher bound-free $M$-shell photoabsorption as that one obtained in the STAR calculations \cite{Krief2018}. Then, the alterations of the bound-free oscillator strengths due to ion correlations that would largely drive the photoabsorption cross-section to higher values \cite{Krief2018}, prove to be significant in the RESEOS calculations only at free electron energies $\eps\lesssim10$ eV. This is illustrated in Fig. {\ref{oscil_Fe_fig}} that presents single-electron oscillator strengths of some bound-free $M$-shell dipole transitions $\alpha \to \beta$:
\begin{displaymath}
f_{\alpha\beta}=\dfrac{(2j_{\beta}+1)}{3(j_{\alpha}+j_{\beta}+1)}\left(\delta_{|j_{\beta}-j_{\alpha}|,1}+\delta_{|j_{\beta}-j_{\alpha}|,0}\,\dfrac{2}{(2l_{\alpha}+1)\,
(2l_{\beta}+1)}\right)\cdot
\end{displaymath}
\begin{equation}\label{oscil_bf}
\cdot\dfrac{r_{\alpha\beta}^{2}}{\eps_{\beta}-\eps_{\alpha}},
\end{equation}
where the radial transition integral $r_{\alpha\beta}$ is taken in the relativistic velocity form \cite{RESEOS2016,Grant1974}:
\begin{displaymath}
r_{\alpha\beta}=c\int\limits_{0}^{\infty}\left((\kap_{\beta}-\kap_{\alpha}+1)\,P_{\alpha}(r)\,Q_{\beta}(r)+\right . 
\end{displaymath}
\begin{equation}\label{matrix_velocity}
\left . +(\kap_{\beta}-\kap_{\alpha}-1)\,Q_{\alpha}(r)\,P_{\beta}(r)\right)dr.
\end{equation}
Here, $c$ is the speed of light, $P_{s}(r)/r$ and $Q_{s}(r)/r$ are the major and the minor radial components of the electron wavefunction, respectively; $\eps_{s}$, $l_{s}$ and $j_{s}$ stand for single-electron energy, orbital and total angular momenta of an atomic subshell $s$, respectively.

These alterations have only a little effect on the photoionization cross-section almost everywhere except for narrow near-threshold regions of photon energies. As a result, RESEOS predicts much smaller increase of the Rosseland mean opacity due to ion correlations (5.5 \%) than the STAR-code calculation \cite{Krief2018} does (20 \%).

It should also be noted that such a modest increase of the Rosseland mean opacity of iron in the RESEOS calculations is contributed not only by the enhancement of the $M$-shell photoionization cross-section, but also by the increased bound-bound absorption at the wings of the $M$-shell transition arrays. At the moment, RESEOS includes the electron-collisional broadening by using the frequency-independent electron-collisional widths, thus overestimating the far-wing photoabsorption of the spectral-line arrays (see, e.g., Ref. \cite{Iglesias2009}). To estimate the role of this effect, additional RESEOS calculations were performed with no electron-collisional broadening included. In these calculations, the account of ion correlations led to the increase of the iron Rosseland mean of only 2.7 \% (from $\kap_{R}^{(IS)}=669.13$ cm$^{2}$/g to $\kap_{R}^{(IC)}=687.08$ cm$^{2}$/g). This increase is actually comprised by the 2.6 \% enhancement of the average $M$-shell occupation number, a minor lowering of the bound-free oscillator strengths at $\eps\gtrsim10$ eV (effectively counterbalancing the first constituent), and a small shift of the photoionization thresholds to higher photon energies mentioned above. The second constituent stems from the fact that the radial integral (\ref{matrix_velocity}) at high free-electron energies is accumulated mostly at small $r$ (where the free-electron wavefunction is devoid of rapid oscillations) for which the electron potentials $V(r)$ (see Fig. {\ref{Potential_STAR_compare_fig} below) are nearly identical regardless of whether they allow for ion correlations or not.
To provide further explanation, one can use
a simple Kramers formula for the cross-section of the bound-free absorption (see, e.g., Refs. \cite{ZRe,IPM2005}) to approximate the relevant oscillator strength utilizing an effective Coulomb potential at $r \simeq r_{IS}, r_{IC}$ with a suitable effective ion-core charge. The latter would be slightly smaller in the IC case than in the IS one due to some ion-correlation enhancement of the average occupation number of the
$M$- and $L$-shell electrons
thus providing a correspondingly increased screening of the nuclear charge $Z$ and therefore lower values of the bound-free oscillator strengths in the IC case at high free-electron energies.

To provide more detailed comparisons with the STAR-code calculations allowing for ion-correlations \cite{Krief2018}, we also calculated the opacity of iron at approximately 2.5 times larger electron density --- at the conditions specific to the convection and radiative zones boundary of the Sun: $T=180$ eV, $\rho=0.473$ g/cm$^{3}$ ($n_{e}=8\cdot10^{22}$ cm$^{-3}$). In Figs. {\ref{gii_STAR_compare_fig}} and {\ref{Potential_STAR_compare_fig}} we present the comparisons of the RESEOS-calculated ion-ion correlation function $g(r)$ and electron potentials $V(r)$ with those ones obtained by using the STAR code \cite{Krief2018}. The RESEOS electron potentials in the ion-correlation calculations were obtained both with and with no regard for the term responsible for the correlations of free electrons and external ions \cite{Starrett2014} that was omitted in the Rozsnyai model\footnote{The Hartree atomic units ($\hbar=m_{e}=e=1$) are used throughout the paper.}:,
\begin{equation}\label{Viec}
V_{Ie}(r)=-\dfrac{n_{i}}{\beta}\int\limits\tilde{C}_{Ie}\left(\left|\vec{r}-\vec{r}_{1}\right|\right)(g(r_{1})-1)\,d\vec{r}_{1}.
\end{equation}
Here, $\beta=1/T$,
\begin{equation}\label{Cie}
\tilde{C}_{Ie}(r)=C_{Ie}(r)-\dfrac{\beta\,\langle Z\rangle}{r},
\end{equation}
where $C_{Ie}(r)$ is the direct electron-ion correlation function \cite{Starrett2013,Starrett2014},
\begin{equation}\label{ionization2}
\langle Z\rangle=n_{e}^{0}/n_{i}
\end{equation}
with $n_{e}^{0}=\lim\limits_{r\to\infty} n_{e}(r)$ being the asymptotic electron density. Unless otherwise stated, the ion-correlation calculations in the present work were performed with the term (\ref{Viec}) included.

One can see from Fig. {\ref{Potential_STAR_compare_fig}} that the ion-sphere potentials obtained with the RESEOS and STAR codes are nearly exactly coincident. The inclusion of ion correlations however shows generally less pronounced modification of the RESEOS-calculated $V(r)$ compared to that one obtained with the STAR code. The form of the RESEOS IC electron potential is governed by the behavior of the relevant ion-ion correlation function demonstrating a smaller overall deviation from the step function than the STAR-code calculated $g(r)$ and maximizing at $r \simeq 1.75r_{0}$ (Fig. {\ref{gii_STAR_compare_fig}}).

At the same time, the STAR $g(r)$ displays a pronounced nonzero values at much smaller distances than the RESEOS $g(r)$ does (see Fig. {\ref{gii_STAR_compare_fig}}). In other words, the STAR code enables external ions to be more closely arranged around the central ion, thus indicating that the effective interionic potential in the Rozsnyai model appears to be more weak than that one in the model of Starrett and Saumon. More close arrangement of external ions, in turn, weakens the screening of the central ion by free electrons. This effect manifests itself through the extension of the electron-potential range (Fig. {\ref{Potential_STAR_compare_fig}}) being clearly longer in the STAR calculation and therefore implying a more pronounced decrease of mean ion charge $Z_{0}$.
It is also seen from the figure that the inclusion of the $V_{Ie}(r)$ term (\ref{Viec}) in the RESEOS calculations extends the electron-potential range as well. However, in the case considered this effect appears to be less pronounced as compared to the sensitivity of the electron potential to the approximation utilized to evaluate ion-ion correlation function.

To examine the similar sensitivity of bound-free oscillator strengths, mean ion charge, and the opacity itself we approximated the STAR-code calculated $g(r)$ by a simple
Debye-H\"{u}ckel-like expression:
\begin{equation}\label{gii_model}
g(r)=\exp(-A\,\exp(-r/B)/r),
\end{equation}
with $A$ and $B$ being the adjustable parameters, and calculated the electron potential with the term (\ref{Viec}) omitted. As expected, the potential obtained fits the STAR-code result rather well (see Fig. {\ref{Potential_STAR_compare_fig}}).

The RESEOS IC calculations done with the approximated ion-ion correlation function (\ref{gii_model}) and the $V_{Ie}(r)$ term (\ref{Viec}) omitted showed that the behavior of the relevant bound-free oscillator strengths is quite similar to that one presented in Fig. \ref{oscil_Fe_fig}: significant enhancement of the bound-free oscillator strengths occurs only at free electron energies $\eps\lesssim10$ eV with a minor lowering of those at all other $\eps\gtrsim 10$ eV.

The values of mean ion charge, average $M$-shell occupation number, and Rosseland mean opacity calculated with various electron potentials are listed in Table {\ref{Fe_180eV_table}}.
One can see that the enhancement of the Rosseland mean opacity is correlated with the increase in the $M$-shell occupancy that occurs due to the extended range of the electron potential (see Fig. {\ref{Potential_STAR_compare_fig}}). Actually, the effect due to higher probabilities of the configurations involving the $M$-shell electrons is largely counterbalanced by the lowering of the oscillator strengths at $\eps\gtrsim10$ eV so that the amplitudes of the monochromatic bound-free $M$-shell opacities in the IS and IC calculations are nearly identical. This is illustrated in Fig. \ref{cross_bf_n3_fig} that presents the relevant RESEOS calculations utilizing the approximation (\ref{gii_model}) to display the effect due to ion correlations in a more observable fashion.

As one can readily see from Fig. {\ref{cross_bf_n3_fig}}, increased monochromatic opacity
$\kap_{bf}(\omega)$ in the IC calculation at $\omega>500$ eV may mostly be attributed to a modest blue shift of the bound-free thresholds stemming from larger binding energies in the potential allowing for ion correlations.

Thus, from Table \ref{Fe_180eV_table} and Fig. \ref{cross_bf_n3_fig} we conclude that all three manifestations of the effect of ion correlations discussed in previous section contribute (with different signs) to the net enhancement of the bound-free $M$-shell opacities observed in the RESEOS calculations utilizing ion-ion pair correlation functions found with both the Rozsnyai and (to lesser extent) the Starrett and Saumon models.

\begin{figure}
 \begin{center}
 \psfrag{xlab}{$\sqrt{r/r_{0}}$}
 \psfrag{ylab}{$g(r)$}
 \includegraphics[scale=0.45,angle=270]{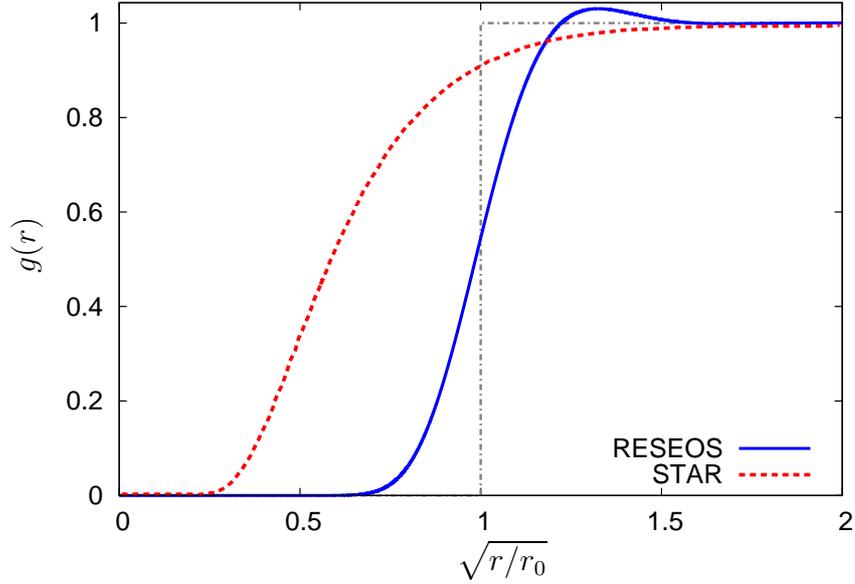}
 \caption{Ion-ion correlation functions for iron at $T=180$ eV, $\rho=0.473$ g/cm$^{3}$ ($n_{e}=8\cdot10^{22}$ cm$^{-3}$) as calculated with the RESEOS (blue solid curve) and STAR \cite{Krief2018} (red dashed curve) codes. Gray dash-dotted curve represents the IS-model pair correlation function.}
 \label{gii_STAR_compare_fig}
 \end{center}
\end{figure}

\begin{figure}
 \begin{center}
 \psfrag{xlab}{$\sqrt{r/r_{0}}$}
 \psfrag{ylab}{$-\sqrt{\max(-V(r),0)/\text{eV}}$}
 \includegraphics[scale=0.45,angle=270]{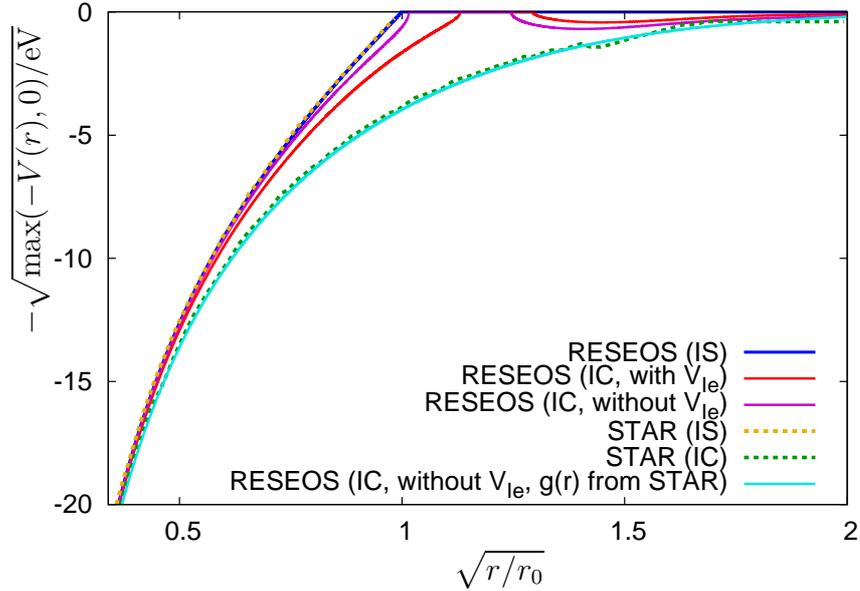}
 \caption{Electron potentials for iron at $T=180$ eV, $\rho=0.473$ g/cm$^{3}$ ($n_{e}=8\cdot10^{22}$ cm$^{-3}$) calculated using the RESEOS code with (red, magenta, and cyan solid curves) and with no (blue solid curve) regard for ion correlations as compared to the corresponding STAR-code calculations \cite{Krief2018} with (green dashed curve) and with no (yellow dashed curve) regard for ion correlations. Red, magenta, and cyan solid curves respectively represent the RESEOS calculations utilizing $V_{Ie}(r)$ (\ref{Viec}), $V_{Ie}(r)\equiv0$, and $V_{Ie}(r)\equiv0$ with the approximation (\ref{gii_model}) of the STAR-code calculated $g(r)$ \cite{Krief2018}.}
 \label{Potential_STAR_compare_fig}
 \end{center}
\end{figure}


\begin{table}
\caption{Mean ion charges $Z_{0}$, average $M$-shell occupation numbers $N_{M}$, and Rosseland mean opacities (disregarding the electron-collisional broadening) $\kap_{R}$ of iron at $T=180$ eV, $\rho=0.473$ g/cm$^{3}$ calculated with RESEOS by using various approximations. $\delta N_{M}$ and $\delta\kap_{R}$ denote the percentage increment of the relevant values due to the effect of ion correlations}\label{Fe_180eV_table}
\begin{tabular}{|c|c|c|c|c|}
\hline
IS/IC&IS&IC&IC&IC\\
model&&&&\\ 
\hline $g(r)$&$\theta(r-r_{0})$&RESEOS&RESEOS&STAR,\\
&&&&Eq. (\ref{gii_model})\\
\hline $V_{Ie}(r)$&$-$&$-$&$+$&$-$\\
\hline $Z_{0}$&15.67&15.64&15.51&14.65\\
\hline $N_{M}$&0.707&0.713&0.729&0.786\\
\hline $\delta N_{M}$&0&0.8 \%&3.1 \%&11.1 \%\\
\hline $\kap_{R}$, $10^{3}$&$1.101$&$1.108$&$1.133$&$1.215$\\
cm$^{2}$/g&&&&\\
\hline $\delta\kap_{R}$&0&0.7 \%&3.0 \%&10.3 \%\\
\hline
\end{tabular}
\end{table}

\begin{figure}
 \begin{center}
 \psfrag{xlab}{$\omega$, eV}
 \psfrag{ylab}{$\kap_{bf}(\omega)$, cm$^{2}$/g}
 \includegraphics[scale=0.45,angle=270]{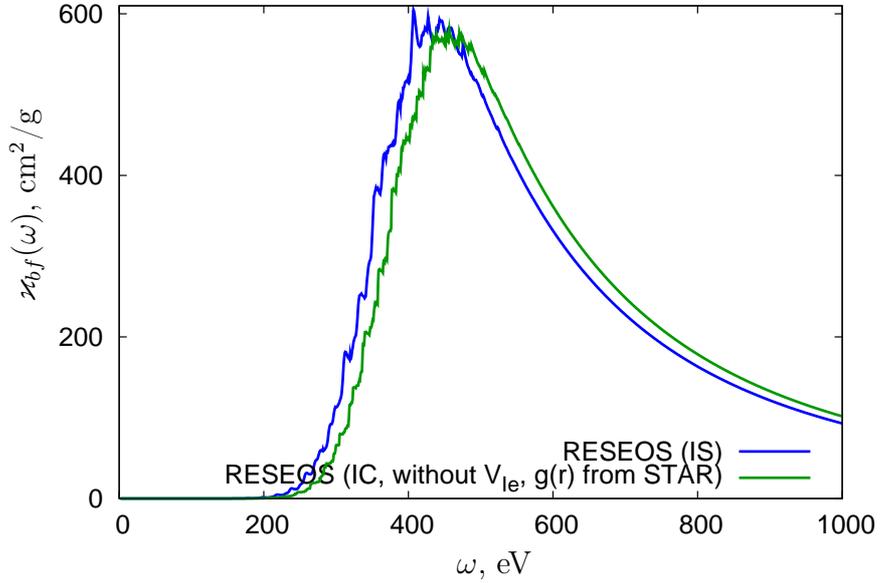}
 \caption{$M$-shell bound-free opacities of iron at $T=180$ eV, $\rho=0.473$ g/cm$^{3}$ calculated by the RESEOS code with no (blue curve) and with (green curve) regard for ion correlations (assuming $V_{Ie}(r)\equiv0$, and utilizing the approximation (\ref{gii_model}) of the STAR-code calculated $g(r)$ \cite{Krief2018})}
 \label{cross_bf_n3_fig}
 \end{center}
\end{figure}

We have also shown (see Table {\ref{Fe_180eV_table}}) that distinctions between these ion-ion pair correlation functions, driven by the difference of the interionic potentials in the ion-correlation models discussed, are responsible for markedly dissimilar enhancement of the plasma opacity due to ion correlations in the relevant RESEOS calculations. The inclusion of the term (\ref{Viec}) in the RESEOS electron potential also contributes to the mean ion charge and opacity, but this contribution only partially counterbalances the distinctions between pair correlation functions. In this connection, we note that structural properties of plasmas calculated by using the Starrett and Saumon model generally agree well \cite{Starrett2013,Starrett2014,Starrett_Daligault_Saumon_2015,Starrett_Saumon_2015} with \textit{ab-initio} calculations and experimental data \cite{Fletcher_2015}, thus providing an appropriate verification of the model.

Next, we compare the interionic potentials in the ion-correlation models of Rozsnyai and Starrett and Saumon under the assumptions of weak non-ideality and weak electron degeneracy --- these assumptions are often valid when the radiation transport is important:
\begin{displaymath}
r_{0}\ll d=\sqrt{\dfrac{T}{4\pi\,Z_{0}\,n_{i}}},\;\;\;r_{0}\ll r_{D}=\sqrt{\dfrac{T}{4\pi\,Z_{0}\,(Z_{0}+1)\,n_{i}}},
\end{displaymath}
\begin{equation}\label{small_nonid}
e^{\beta\,\mu_{e}}\ll1,
\end{equation}
where $d$ and $r_{D}$ are the electron and the total Debye radii, respectively, $\mu_{e}$ is the electron chemical potential. It may be shown (see Appendix A) that in the limit (\ref{small_nonid}) the structural properties of plasmas in the Starrett and Saumon model coincide with those ones obtained by using the Debye-H\"{u}ckel model.
Specifically, the pair interionic potential in the limit (\ref{small_nonid}) takes the form:
\begin{equation}\label{Vii_DH}
V_{II}(r)=\dfrac{Z_{0}^{2}}{r}\,e^{-r/d},
\end{equation}
while the self-consistent electron potential
\begin{equation}\label{V_DH}
V(r)=-\dfrac{Z_{0}}{r}\,e^{-r/r_{D}}.
\end{equation}

If one disregards the electron exchange and correlation effects, the interionic potential in the Rozsnyai model \cite{Rozsnyai1991} is written as
\begin{equation}\label{Vii_Rozsnyai}
\tilde{V}_{II}(r)=r\,V^{2}(r).
\end{equation}
Next, we consider the asymptotic case (\ref{small_nonid}) assuming that the potential $V(r)$ is determined by Eq. (\ref{V_DH}).\footnote{Though the asymptotic expression for the electron potential in the Rozsnyai model may differ from that one given by Eq. (\ref{V_DH}), at the moment we are interested only in the relation between potentials $V_{II}(r)$ and $\tilde{V}_{II}(r)$ for the same potential $V(r)$.} Then the Rozsnyai model yields:
\begin{equation}\label{Vii_Rozsnyai_asympt}
\tilde{V}_{II}(r)=\dfrac{Z_{0}^{2}}{r}\,e^{-2r/r_{D}},
\end{equation}
while the ratio of the interionic potentials in the Rozsnyai and the Starrett and Saumon models becomes
\begin{equation}\label{Vii_ratio}
\dfrac{\tilde{V}_{II}(r)}{V_{II}(r)}=e^{-r/r_{D}}\cdot e^{-r\left(1/r_{D}-1/d\right)}.
\end{equation}
The first and the second factors in the right-hand side of Eq. (\ref{Vii_ratio}) are both less than unity. Hence, it is clear that the interionic potential in the Rozsnyai model is indeed weaker
than that one in the model of Starrett and Saumon.

We also note that later Rozsnyai employed some other expression for the interionic potential \cite{Rozsnyai2014}:
\begin{equation}\label{Vii_Rozsnyai_alt}
\tilde{V}_{II}(r)=\dfrac{1}{r}\left(Z-4\pi\int\limits_{0}^{r}\left(n_{e}(r_{1})-n_{e}^{0}\,g(r_{1})\right)r_{1}^{2}\,dr_{1}\right)^{2}
\end{equation}
yielding the ratio of the interionic potentials under the assumptions (\ref{small_nonid}) in the form:
\begin{equation}\label{Vii_ratio_alt}
\dfrac{\tilde{V}_{II}(r)}{V_{II}(r)}=e^{-r/r_{D}}\cdot e^{-r\left(1/r_{D}-1/d\right)}\left(1+\dfrac{r}{r_{D}}\right)^{2}.
\end{equation}
Though the ratio (\ref{Vii_ratio_alt}) is not always less than unity, this is unlikely to be important here since, to the best of our knowledge, the potential (\ref{Vii_Rozsnyai_alt}) was not used in the opacity calculations based on the Rozsnyai model \cite{Rozsnyai1991,Rozsnyai1992,Krief2018}.

\begin{figure}
 \begin{center}
 \psfrag{xlab}{$\omega$, eV}
 \psfrag{ylab}{$\kap(\omega)$, cm$^{2}$/g}
 \includegraphics[scale=0.45,angle=270]{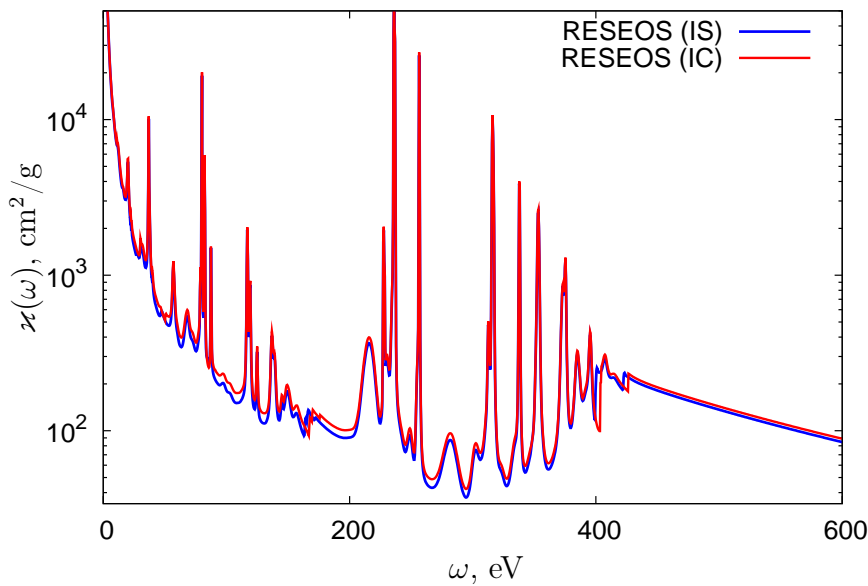}
 \caption{The monochromatic opacity of aluminum at a temperature $T=100$ eV and material density $\rho=0.0045$ g/cm$^{3}$ calculated using the RESEOS code with (red curve, $\kap_{R}=78.73$ cm$^{2}$/g) and with no (blue curve, $\kap_{R}=74.06$ cm$^{2}$/g) regard for ion correlations.}
 \label{Spectr_Al_IS_IC_fig}
 \end{center}
\end{figure}

\begin{figure}
 \begin{center}
 \psfrag{xlab}{$r/r_{0}$}
 \psfrag{ylab}{$g(r)$}
 \includegraphics[scale=0.45,angle=270]{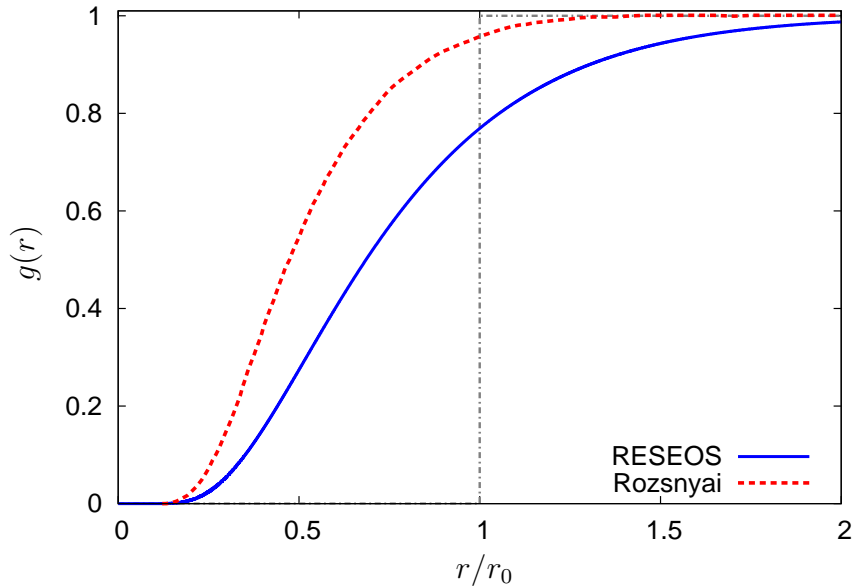}
 \caption{Ion-ion correlation functions for aluminum at $T=100$ eV, $\rho=0.0045$ g/cm$^{3}$ as calculated by using RESEOS (blue solid curve) and by Rozsnyai \cite{Rozsnyai1991} (red dashed curve). Gray dash-dotted curve represents the IS-model pair correlation function.}
 \label{gii_Al_fig}
 \end{center}
\end{figure}

Then we consider the case of aluminum at a temperature $T=100$ eV and material density $\rho=0.0045$ g/cm$^{3}$. The monochromatic opacity calculated by using both the IS and IC options of RESEOS are presented in Fig. {\ref{Spectr_Al_IS_IC_fig}}. The use of the IS model showed that the value of mean ion charge ($Z_{0}=10.77$) is nearly coincident here with the similar result of Ref. \cite{Rozsnyai1991} ($Z_{0}=10.76$). If the ion-correlation lowering of mean ion charge with the Starrett and Saumon model were as large as that one with the Rozsnyai model (to $Z_{0}=10.63$), then the value of $\kap_{R}$ would likely be increased in a similar way --- by a factor of $\simeq 1.5$. The use of the Starrett and Saumon model however yields $Z_{0}=10.73$ being quite close to that one in the IS model. As a result, according to our calculations ion-correlation enhancement of the Rosseland mean opacity appears to be of only 6 \% instead of 55 \% in Ref. \cite{Rozsnyai1991}.

The aluminum ion-ion correlation functions calculated using the Rozsnyai and the Starrett and Saumon models show similar distinctions as those ones in the case of iron above (Fig. {\ref{gii_Al_fig}}). However, these distinctions provide no predominant contribution to the difference of mean ion charges. To illustrate this, we performed the IC RESEOS calculation using the $g(r)$ function of Rozsnyai \cite{Rozsnyai1991} approximated with Eq. (\ref{gii_model}) and omitting the term (\ref{Viec}) in the electron potential. This calculation yielded the mean ion charge ($Z_{0}=10.72$) and the enhancement of the Rosseland mean opacity (of only 10 \%) being, respectively, almost coincident or quite comparable with the relevant values obtained by using the Starrett and Saumon model.

We should also note one more issue apparently being important in the aluminum case considered. This issue is associated with the choice of the electron chemical potential $\mu_{e}$. Following Starrett and Saumon \cite{Starrett2013,Starrett2014}, the chemical potential in the ion-correlation RESEOS calculations was assumed to be the same as in the relevant NWS calculations. In this approximation, the ion-correlation effect on mean ion charge manifests itself only through the shifts of bound-electron energies due to the extension of the electron-potential range.

We emphasize that the evaluation of the ion-correlation correction to the chemical potential seems to be rather problematic issue for the Starrett and Saumon model. For example, an application of the variational principle for $\mu_{e}$ \cite{Chihara2016} in the framework of the Starrett and Saumon model leads to nonphysical results (see Appendix A). Therefore, we restrict ourselves to the use of the chemical potential found in the NWS model from the charge-conservation condition inside the atomic cell.

Unlike the Starrett and Saumon model, the chemical potential in the Rozsnyai model is modified due to the effect of ion correlations. Since no specific condition to evaluate the chemical potential in the Rozsnyai model is provided, we can only assume that the chemical potential is matched to ensure the charge conservation inside some large but finite volume --- in the  single-center model the chemical potential cannot be found from the charge conservation condition inside an infinite volume \cite{Blenski2007_VAAQP}. The account of ion correlations in the Rozsnyai model generally leads to the decrease of $\mu_{e}$ \cite{Rozsnyai1991}, thus being able to significantly contribute to the decrease of $Z_{0}$ and the corresponding increase of opacity.

However, $Z_{0}^{(IC)}=10.63$, as predicted by Rozsnyai \cite{Rozsnyai1991} for aluminum at $T=100$ eV, $\rho=0.0045$ g/cm$^{3}$, appears to be farther apart from the relevant values given by two chemical-picture-based models --- ChemEOS \cite{Kilcrease2015} and CP-SC \cite{Loboda2009}, being quite appropriate for the low-density case considered, --- than the value $Z_{0}^{(IC)}=10.73$ obtained by using the Starrett and Saumon model. ChemEOS and CP-SC adopt the approximations of the Monte-Carlo and hypernetted-chain calculated data on the interparticle Coulomb contributions to the Helmholtz free energy \cite{Chabrier1998,Potekhin2000} and therefore consistently allow for ion correlations. These models both predict nearly the same mean ion charges: $Z_{0}=10.84$ (ChemEOS) \cite{ATOMIC} and $Z_{0}=10.83$ (CP-SC) being markedly closer to the RESEOS result than to that one predicted by Rozsnyai \cite{Rozsnyai1991}. Since the distinctions in the ion-ion correlation functions calculated using the Rozsnyai and the Starrett and Saumon models provide only insignificant contribution to the difference of the relevant mean ion charges, this difference should be attributed to the effect of the ion-correlation modification of the chemical potential in the Rozsnyai model. As this effect increases the departure of mean ion charge from the values provided by the ChemEOS and CP-SC models, it is likely that a method to evaluate the chemical potential in the Rozsnyai model with the account of ion correlations needs some revision.

\begin{figure}
 \begin{center}
 \psfrag{xlab}{$\omega$, eV}
 \psfrag{ylab}{$\kap(\omega)$, cm$^{2}$/g}
 \includegraphics[scale=0.45,angle=270]{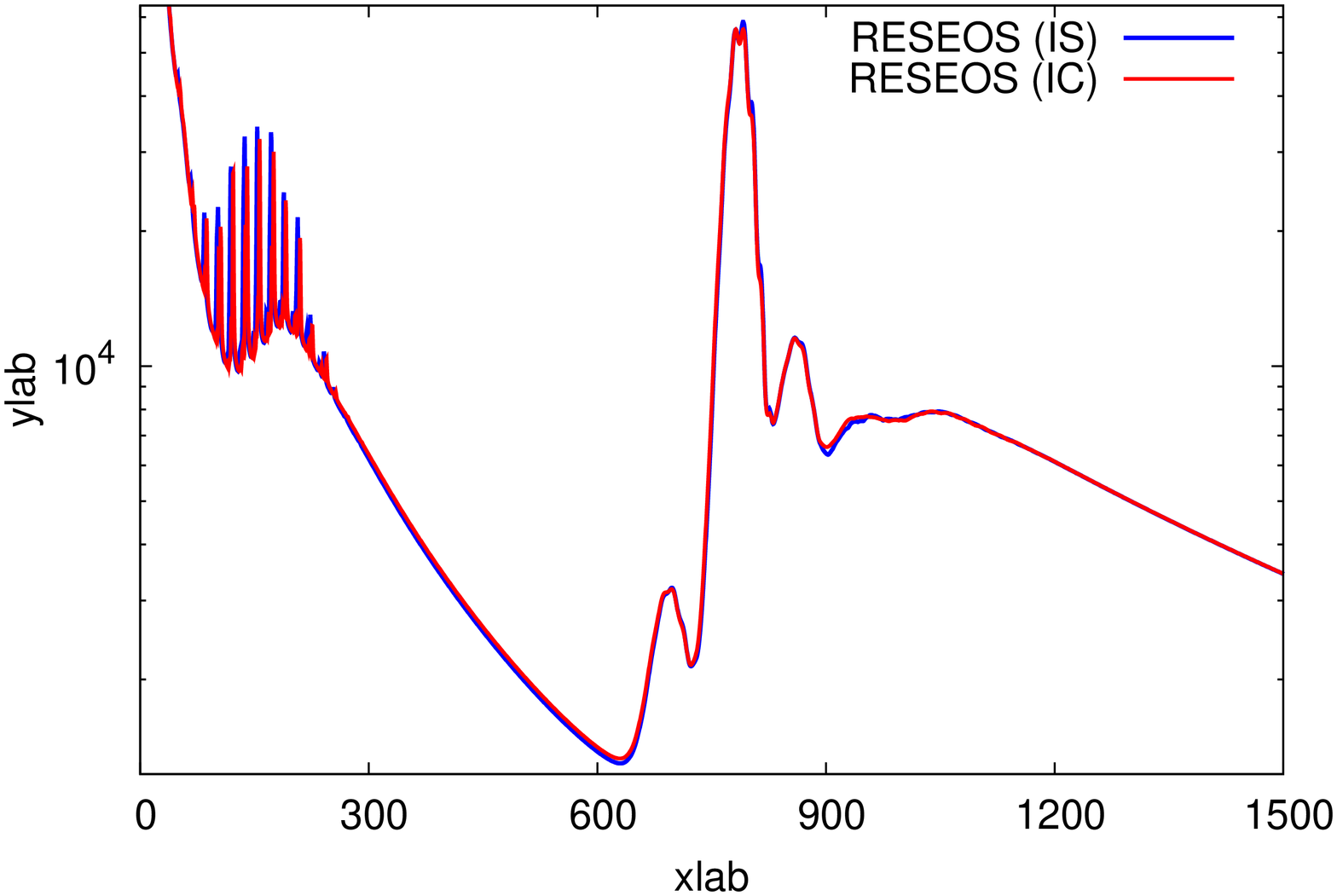}
 \caption{The monochromatic opacity of iron at a temperature $T=125$ eV and material density $\rho=4.46$ g/cm$^{3}$ calculated using the RESEOS code with (red curve, $\kap_{R}=3.21\cdot10^{3}$ cm$^{2}$/g) and with no (blue curve, $\kap_{R}=3.15\cdot10^{3}$ cm$^{2}$/g) regard for ion correlations.}
 \label{Spectr_Fe_hd_IS_IC_fig}
 \end{center}
\end{figure}

\begin{figure}
 \begin{center}
 \psfrag{xlab}{$\omega$, eV}
 \psfrag{ylab}{$\kap(\omega)$, cm$^{2}$/g}
 \includegraphics[scale=0.45,angle=270]{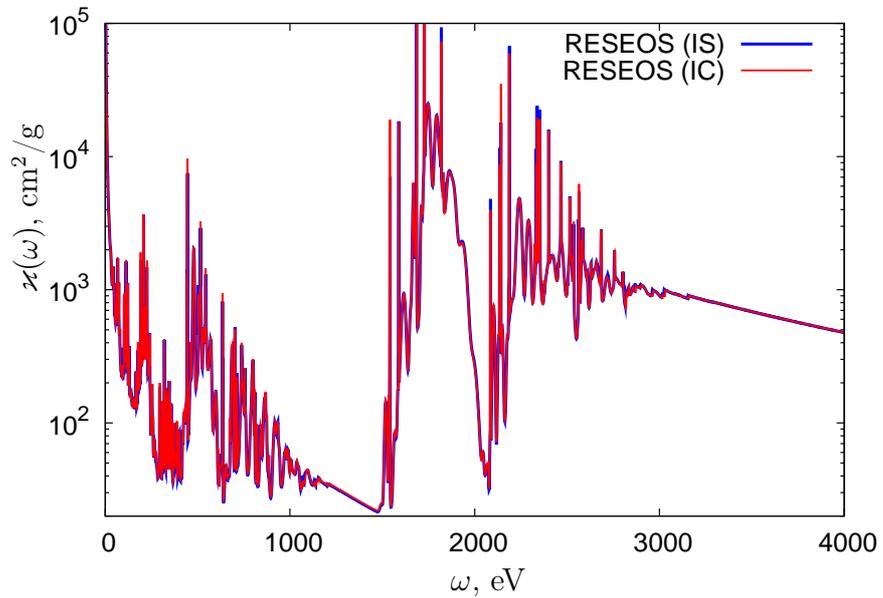}
 \caption{The monochromatic opacity of bromine at a temperature $T=270$ eV and material density $\rho=0.026$ g/cm$^{3}$ calculated using the RESEOS code with (red curve, $\kap_{R}=60.17$ cm$^{2}$/g) and with no (blue curve, $\kap_{R}=59.62$ cm$^{2}$/g) regard for ion correlations.}
 \label{Spectr_Br_IS_IC_fig}
 \end{center}
\end{figure}

Finally, we consider two more, higher-density, cases previously analyzed in Ref. \cite{Rozsnyai1991}: iron at $T=125$ eV, $\rho=4.46$ g/cm$^{3}$ (Fig. {\ref{Spectr_Fe_hd_IS_IC_fig}}) and bromine at $T=270$ eV, $\rho=0.026$ g/cm$^{3}$ (Fig. {\ref{Spectr_Br_IS_IC_fig}}). For these cases, a perceptible decrease of mean ion charges (by 18 and 16 \% for iron and bromine, respectively) and the relevant enhancement of the Rosseland means (by 14 and 11 \% for iron and bromine, respectively) due to the ion-correlation effect were predicted \cite{Rozsnyai1991}. Contrary to these predictions, and much as in the cases considered above, we obtained only a minor decrease of mean ion charges (by 1 and 0.1 \% for iron and bromine, respectively) and the enhancement of the Rosseland means (by 2 and 1 \% for iron and bromine, respectively). At the same time, we obtained a pronounced decrease of the mean ion charge (by 11 \%) and the relevant enhancement of the Rosseland mean opacity (by 15 \%) for the iron case when using the approximation (\ref{gii_model}) of the Rozsnyai-model $g(r)$ \cite{Rozsnyai1991}. Therefore, in this case a disagreement of the results obtained with the use of the Starrett and Saumon model and the results of Ref. \cite{Rozsnyai1991} seems to be mostly driven by the distinctions in the ion-ion correlation functions rather than in the chemical potentials.

\section{Conclusion}

The calculations we performed using the Starrett and Saumon model show the smaller effect of ion correlations on plasma opacities than that one found in earlier work \cite{Rozsnyai1991,Rozsnyai1992,Krief2018} basing on the Rozsnyai model. We have shown that the Rozsnyai model overestimates the effect of ion correlations due to the underestimation of ion-ion repulsion. At the same time, a more accurate method to calculate the electron chemical potential is required to consistently evaluate the effect of ion correlations on plasma opacities with the model of Starrett and Saumon. If the method to evaluate the chemical potential in the up-to-date version of the Starrett and Saumon model does not lead to significant inaccuracies, the account of ion correlations appears to be not very important in the context of the solar opacity problem, thus favoring the assumption that this problem is caused by the existence of some unaccounted deficiencies of the solar modeling (see, e.g., Ref. \cite{Pain2017}) rather than by any substantial inaccuracies in the theoretical opacity data.

\section*{Appendix A. Starrett and Saumon model under the conditions of weak non-ideality and weak electron degeneracy}
\setcounter{section}{1}%
\setcounter{equation}{0}%
\setcounter{figure}{0}%
\setcounter{table}{0}%
\renewcommand{\thesection}{A}

Let us now consider the equations of the Starrett and Saumon model assuming the smallness of the electron-ion coupling parameter:
\begin{equation}\label{gamma_ei_small}
\Gamma_{ei}=\dfrac{\beta\,Z_{0}}{r_{0}}\ll1
\end{equation}
(the case of high temperatures and/or low material densities). Here, mean ion charge $Z_{0}$ is defined as the difference of nuclear charge $Z$ and mean number of bound electrons that belong to a single atom:
\begin{equation}\label{ionization1}
Z_{0}=Z-4\pi\int\limits_{0}^{\infty}r^{2}\,n_{e,b}(r)\,dr,
\end{equation}
where $n_{e,b}(r)$ is the bound-electron density.

Under the condition (\ref{gamma_ei_small}), one can neglect the size of ion core (pointlike-ion approximation):
\begin{equation}\label{dens_bound_delta}
n_{e,b}(r)\approx(Z-Z_{0})\,\delta\left(\vec{r}\right).
\end{equation}

We also neglect free-electron relativistic corrections, since the temperatures we consider are much less than $m_{e}\,c^{2}\approx511$ keV. Mean ion charge is assumed to be large enough,
\begin{equation}\label{strong_ioinization}
Z_{0}\gg1,
\end{equation}
to take no account of the exchange-correlation potential $V_{xc}\sim\left(n_{e}^{0}\right)^{1/3}\sim Z_{0}^{1/3}/r_{0}$ as compared to the electrostatic one: $V_{el}\sim Z_{0}/r_{0}$.

Then, we omit the term (\ref{Viec}) in the electron potential responsible for the correlations of free electrons and external ions \cite{Starrett2013,Starrett2014}. The possibility to do that under the conditions considered will be shown below.

Besides that, we use the semiclassical approximation for free electrons as it works well at high temperatures.

With due regard for the assumptions above, the electron potential $V(r)$ at $r>0$ becomes
\begin{equation}\label{potential_el}
V(r)=-\dfrac{Z_{0}}{r}+\int\limits\dfrac{d\vec{r}_{1}\left(n_{e}(r_{1})-n_{e}^{0}\,g(r_{1})\right)}{\left|\vec{r}-\vec{r}_{1}\right|},
\end{equation}
where
\begin{equation}\label{elec_dens}
n_{e}(r)=\dfrac{\sqrt{2}}{\pi^{2}\,\beta^{3/2}}\,I_{1/2}\left(\beta\,(\mu_{e}-V(r))\right)
\end{equation}
with $I_{k}(x)=\int\limits_{0}^{\infty} \dfrac{y^{k}\,dy}{1+\exp(y-x)}$ being the Fermi-Dirac integral.

Under the assumption (\ref{gamma_ei_small}), characteristic values of the potential are small as compared to the temperature, thus enabling one to linearize Eq. (\ref{elec_dens}):
\begin{equation}\label{dens_expansion}
n_{e}(r)\approx n_{e}^{0}-\dfrac{V(r)}{4\pi\,d^{2}},
\end{equation}
where $d$ is the electron Debye radius with the account of electron degeneracy \cite{Ebeling1991}:
\begin{equation}\label{Debye}
\dfrac{1}{d^{2}}=2\pi\,\beta\,n_{e}^{0}\,\dfrac{I_{-1/2}(\beta\,\mu_{e})}{I_{1/2}(\beta\,\mu_{e})}.
\end{equation}
Substituting Eq. (\ref{dens_expansion}) into Eq. (\ref{potential_el}) and using the Fourier transformation one gets:
\begin{displaymath}
V(q)=-\dfrac{4\pi\left(Z_{0}+n_{e}^{0}\,h(q)\right)}{q^{2}+1/d^{2}}\Rightarrow V(r)=-\dfrac{Z_{0}}{r}\,e^{-r/d}-
\end{displaymath}
\begin{equation}\label{V_Fourier}
-n_{e}^{0}\int\limits d\vec{r}_{1}
\left(g\left(\left|\vec{r}-\vec{r}_{1}\right|\right)-1\right)\dfrac{e^{-r_{1}/d}}{r_{1}},
\end{equation}
where $h(r)=g(r)-1$.
Hereafter we use the notations:
\begin{displaymath}
f(q)=\dfrac{4\pi}{q}\int\limits_{0}^{\infty}r\,f(r)\,\sin(q\,r)\,dr,
\end{displaymath}
\begin{equation}\label{Fourier_dir_inv}
f(r)=\dfrac{1}{2\pi^{2}\,r}\int\limits_{0}^{\infty} q\,f(q)\,\sin(q\,r)\,dq.
\end{equation}

To this point, no assumptions have been made that would specify the form of the correlation function $g(r)$. Therefore, one can substitute, e.g., the step function (\ref{gii_0}) into Eq. (\ref{V_Fourier}), thus yielding
\begin{equation}\label{V_IS_ht}
V(r)=\left\{\begin{array}[c]{r}-\dfrac{Z_{0}}{r}\,e^{-r/d}+\dfrac{1}{2\beta\,r}\left((r_{0}+d)\left(e^{-(r_{0}+r)/d}-
e^{-(r_{0}-r)/d}\right)+2r\right)\\ \text{if }r\leqslant r_{0},\\
-\dfrac{Z_{0}}{r}\,e^{-r/d}+\dfrac{1}{2\beta\,r}\,e^{-r/d}\left((r_{0}+d)\,e^{-r_{0}/d}+(r_{0}-d)\,e^{r_{0}/d}\right)
\\ \text{if }r\geqslant r_{0}.\end{array}\right .
\end{equation}
By neglecting the corrections of the order of $\Gamma_{ei}$ in Eq. (\ref{V_IS_ht}) we get the well-known expression for the electrostatic potential in the approximation of uniform free electron density in the ion-sphere-restricted (IS) model:
\begin{equation}\label{V_dens_const}
V(r)=-\dfrac{Z_{0}}{r}\left(1-\dfrac{3r}{2r_{0}}+\frac{1}{2}\left(\dfrac{r}{r_{0}}\right)^{3}\right)\theta(r_{0}-r).
\end{equation}

To obtain a realistic pair correlation function and the relevant electron potential, we additionally assume the smallness of the ion-ion coupling parameter:
\begin{equation}\label{gamma_ii_small}
\Gamma_{ii}=\dfrac{\beta\,Z_{0}^{2}}{r_{0}}\ll1,
\end{equation}
and the smallness of the electron degeneracy parameter:
\begin{equation}\label{degen_small}
\frac{1}{2}\beta\left(3\pi^{2}\,n_{e}^{0}\right)^{\frac{2}{3}}\ll1\Rightarrow e^{\beta\,\mu_{e}}\ll1.
\end{equation}
In this approximation, one can simplify the expression (\ref{Debye}) for the electron Debye radius:
\begin{equation}\label{Debye_nondeg}
\dfrac{1}{d^{2}}=4\pi\,\beta\,n_{e}^{0}.
\end{equation}
Then, we also take no account of the difference of mean ion charges as defined by Eqs. (\ref{ionization1}) and (\ref{ionization2}) --- it may be shown that in the semiclassical approximation $\left(Z_{0}-\langle Z\rangle\right)/\langle Z\rangle=O\left(\Gamma_{ei}\right)$ at least in the IS case.

In the Starrett and Saumon model, the function $g(r)$ is the solution of the Ornstein-Zernike equation \cite{Chihara1991}:
\begin{equation}\label{OrnZern}
h(q)=C(q)+n_{i}\,h(q)\,C(q)
\end{equation}
with the hypernetted-chain closure relation\footnote{The alternative closure relations \cite{Percus_Yevick_1958,Martynov1983} do not provide the Debye-H\"{u}ckel limit.}
\begin{equation}\label{hyper_closer}
1+h(r)=\exp\left(-\beta\,V_{II}(r)+h(r)-C(r)\right),
\end{equation}
where $C(r)$ is the direct correlation function, $V_{II}(r)$ is the effective pair interionic potential allowing for the screening of ions by free electrons. The Fourier transform of $V_{II}(r)$ is expressed through the Fourier transform of the density $n_{e}^{scr}(r)$ of free electrons screening an ion \cite{Starrett2013}:
\begin{equation}\label{Vii}
V_{II}(q)=\dfrac{4\pi\,Z_{0}^{2}}{q^{2}}-\dfrac{1}{\beta}\,C_{Ie}(q)\,n_{e}^{scr}(q).
\end{equation}
Here,
\begin{equation}\label{Cie_Fourier}
C_{Ie}(q)=-\beta\,n_{e}^{scr}(q)\left(1+\chi_{ee}^{0}(q)\,C_{ee}(q)/\beta\right)/\chi_{ee}^{0}(q)
\end{equation}
(see Ref. \cite{Starrett2013}), where $C_{ee}(r)$ is the direct electron-electron correlation function, $\chi_{ee}^{0}(q)$ is the response function of non-interacting electrons that reduces to the product $-n_{e}^{0}\,\beta$ in the limit of weak electron degeneracy considered \cite{Starrett2013}.

The screening electron density in the Starrett and Saumon model is set equal to the pseudoatom free electron density \cite{Starrett2013,Starrett2014}:
\begin{equation}\label{screening}
n_{e}^{scr}(r)=n_{e}^{PA}(r)-n_{e,b}(r)
\end{equation}
with the pseudoatom electron density $n_{e}^{PA}(r)$ being defined as the difference of the electron densities $n_{e}(r)$ and $n_{e}^{ext}(r)$ in the full and external systems, respectively. The external system is a system without central nucleus ($Z=0$) and with the same chemical potential $\mu_{e}$ and pair correlation function $g(r)$ as in the full system. Eq. (\ref{screening}) connects the system of Eqs. (\ref{OrnZern}) -- (\ref{Cie_Fourier}) with the electron density (\ref{elec_dens}) and the potential (\ref{potential_el}).

Taking into account Eqs. (\ref{dens_bound_delta}) and (\ref{screening}), one gets $n_{e}^{ext}(r)$ in a form similar to Eq. (\ref{dens_expansion}), thus leading to the following expressions:
\begin{displaymath}
n_{e}^{ext}(r)\approx n_{e}^{0}-\dfrac{V^{ext}(r)}{4\pi\,d^{2}},\;\;n_{e}^{scr}(r)\approx-\dfrac{V(r)-V^{ext}(r)}{4\pi\,d^{2}}\equiv-\dfrac{V^{PA}(r)}{4\pi\,d^{2}},
\end{displaymath}
\begin{displaymath}
V^{ext}(r)\approx-n_{e}^{0}\int\limits d\vec{r}_{1}\left(g\left(\left|\vec{r}-\vec{r}_{1}\right|\right)-1\right)\dfrac{e^{-r_{1}/d}}{r_{1}}\Rightarrow V^{PA}(r)\approx,
\end{displaymath}
\begin{equation}\label{screening_res}
\approx-\dfrac{Z_{0}}{r}\,e^{-r/d},\;\;n_{e}^{scr}(r)\approx\dfrac{Z_{0}}{4\pi\,d^{2}\,r}\,e^{-r/d},\;\;n_{e}^{scr}(q)=\dfrac{Z_{0}}{1+\left( q\,d\right)^{2}}.
\end{equation}

The Fourier transform of the direct electron-electron correlation function $C_{ee}(q)$ may be expressed in terms of the local field correction $G_{ee}(q)$ \cite{Chihara1999}:
\begin{equation}\label{Cee_Fourier}
C_{ee}(q)=-\dfrac{4\pi\beta}{q^{2}}\left(1-G_{ee}(q)\right).
\end{equation}
In what follows, we are mostly interested in the expression for the pair correlation function $g(r)$ at large radii $r\gtrsim r_{0}$ corresponding to small $q\lesssim1/r_{0}$. At small $q$ $\left(q\ll\left(n_{e}^{0}\right)^{1/3}\sim Z_{0}^{1/3}/r_{0}\right)$ the local field correction becomes \cite{Ichimaru1981}
\begin{equation}\label{Gee_small_q}
G_{ee}(q)\approx\gamma_{0}\,\dfrac{q^{2}}{\left(3\pi^{2}\,n_{e}^{0}\right)^{2/3}},
\end{equation}
where $\gamma_{0}$ is the coefficient of the order of unity connected with the compressibility of electron gas. Therefore,
\begin{displaymath}
1+\chi_{ee}^{0}(q)\,C_{ee}(q)/\beta\approx1+\dfrac{4\pi\,\beta\,n_{e}^{0}}{q^{2}}\left(1-\dfrac{\gamma_{0}\,q^{2}}{\left(3\pi^{2}\,n_{e}^{0}\right)^{2/3}}\right) =1+\dfrac{1}{\left(q\,d\right)^{2}}-
\end{displaymath}
\begin{equation}\label{Cee_term}
-\gamma_{0}\left(\dfrac{16}{3\pi^{2}\,Z_{0}^{2}}\right)^{1/3}\Gamma_{ei}\approx1+\dfrac{1}{\left(q\,d\right)^{2}},
\end{equation}
which is to say that the local field correction in Eq. (\ref{Cee_Fourier}) may be neglected here under the conditions (\ref{gamma_ei_small}), (\ref{gamma_ii_small}), and (\ref{degen_small}) considered.

The Fourier transform of the direct electron-ion correlation function (\ref{Cie_Fourier}) thus becomes
\begin{equation}\label{Cie_Fourier_res}
C_{Ie}(q)=\dfrac{n_{e}^{scr}(q)}{n_{e}^{0}}\left(1+\dfrac{1}{\left(q\,d\right)^{2}}\right)=\dfrac{4\pi\,\beta\,Z_{0}}{q^{2}}.
\end{equation}
On the other hand, according to Eq. (\ref{Cie}) the general expression for the function $C_{Ie}(q)$ is:
\begin{equation}\label{Cie_Fourier_common}
C_{Ie}(q)=\dfrac{4\pi\,\beta\,Z_{0}}{q^{2}}+\tilde{C}_{Ie}(q).
\end{equation}
The comparison of Eqs. (\ref{Cie_Fourier_res}) and (\ref{Cie_Fourier_common}) shows that the correction $\tilde{C}_{Ie}(r)$ may also be neglected under the conditions (\ref{gamma_ei_small}), (\ref{gamma_ii_small}), and (\ref{degen_small}) and the term (\ref{Viec}) in the electron potential may therefore be omitted, as stated above.

For the effective interionic potential (\ref{Vii}) one gets
\begin{equation}\label{Vii_res}
V_{II}(q)=\dfrac{4\pi\,Z_{0}^{2}}{q^{2}+1/d^{2}}\Rightarrow V_{II}(r)=\dfrac{Z_{0}^{2}}{r}\,e^{-r/d}.
\end{equation}

Taking into account Eq. (\ref{gamma_ii_small}), one gets $V_{II}(r)$ at $r\gtrsim r_{0}$:
\begin{equation}\label{Vii_small}
\beta\,V_{II}(r)\lesssim\dfrac{\beta\,Z_{0}^{2}}{r_{0}}\ll1.
\end{equation}
The values of $h(r)$ and $C(r)$ are also small as compared to unity. Therefore, one can expand the exponent in Eq. (\ref{hyper_closer}) and obtain
\begin{equation}\label{C_res}
C(r)\approx-\beta\,V_{II}(r)=-\dfrac{\beta\,Z_{0}^{2}}{r}\,e^{-r/d}\Rightarrow C(q)=-\dfrac{4\pi\,\beta\,Z_{0}^{2}}{q^{2}+1/d^{2}}.
\end{equation}
With Eqs. (\ref{OrnZern}) and (\ref{C_res}) one gets
\begin{equation}\label{h_res}
h(q)=-\dfrac{4\pi\,\beta\,Z_{0}^{2}}{q^{2}+1/r_{D}^{2}}\Rightarrow g(r)=1-\dfrac{\beta\,Z_{0}^{2}}{r}\,e^{-r/r_{D}}.
\end{equation}
Thus, the ion-ion pair correlation function is expressed in terms of the total Debye radius $r_{D}$ (Eq. (\ref{small_nonid})) while the pair interionic potential is dependent on the electron Debye radius.

Substituting Eq. (\ref{h_res}) into Eq. (\ref{V_Fourier}), one gets
\begin{equation}\label{V_res}
V(r)=-\dfrac{Z_{0}}{r}\,e^{-r/r_{D}}.
\end{equation}
It follows from Eqs. (\ref{Vii_res}), (\ref{h_res}), and (\ref{V_res}) that the interparticle potentials and pair correlation functions obtained with the Starrett and Saumon model coincide with those ones obtained with the Debye-H\"{u}ckel model in the limit of weak non-ideality and weak electron degeneracy.

For this limit we then consider the evaluation of the chemical potential. The condition for the chemical potential in the average-atom model based on the variational principle and the cluster expansion \cite{Blenski2007_VAAQP,Piron2011_VAAQP} of the number of electrons and the electron free energy follows from the minimum condition for the grand thermodynamic potential relative to variations of the asymptotic free-electron density $n_{e}^{0}$. Such condition was derived in Ref. \cite{Chihara2016}:
\begin{equation}\label{chem_cond}
4\pi\int\limits_{0}^{\infty} r^{2}\,V_{el}(r)\,g(r)\,dr=\dfrac{\mu_{e}}{n_{i}}\left(1+4\pi\,n_{i}\int\limits_{0}^{\infty} r^{2}(g(r)-1)\,dr\right),
\end{equation}
where $V_{el}$ is the classical electrostatic potential --- the total potential without exchange-correlation contributions (in the limit considered $V_{el}(r)=V(r)$). Substituting the potential (\ref{V_res}) and the pair correlation function (\ref{h_res}) into the left-hand side of Eq. (\ref{chem_cond}) and omitting the small correction of the order of $\Gamma_{ii}$, one gets
\begin{equation}\label{chem_cond_lhs}
-\dfrac{1}{\beta\,(Z_{0}+1)\,n_{i}}.
\end{equation}
At the same time, the substitution of the pair correlation function (\ref{h_res}) into the right-hand side of Eq. (\ref{chem_cond}) yields
\begin{equation}\label{chem_cond_rhs}
\dfrac{\mu_{e}}{(Z_{0}+1)\,n_{i}}.
\end{equation}
The expression (\ref{chem_cond_rhs}) does not coincide with Eq. (\ref{chem_cond_lhs}) and takes considerably larger absolute values at pretty high temperatures. It means that the Starrett and Saumon model would not coincide with the Debye-H\"{u}ckel model in the high temperature/low density limit if the chemical potential were found from Eq. (\ref{chem_cond}).\footnote{In practice, the chemical potential in the model of Starrett and Saumon is set equal to the chemical potential in the NWS model.} This drawback may result from the use of the first-order cluster expansion for the number of electrons,
\begin{equation}\label{cluster_number}
N_{e}=\dfrac{n_{e}^{0}}{n_{i}}+\int\limits\left(n_{e}(r)-n_{e}^{0}\right)d\vec{r}=Z+n_{e}^{0}\int\limits(g(r)-\theta(r-r_{0}))\,d\vec{r},
\end{equation}
and the electron Helmholtz free energy
\begin{equation}\label{cluster_energy}
F_{e}=\dfrac{f_{e}^{0}}{n_{i}}+\int\limits\left(f_{e}(r)-f_{e}^{0}\right)d\vec{r}
\end{equation}
per a single atom, where $f_{e}^{0}$ $\left(f_{e}(r)\right)$ represents the free-energy density of the homogeneous (inhomogeneous) electron gas. From Eq. (\ref{cluster_number}) one can see that in the ion-correlation case $(g(r)\neq\theta(r-r_{0}))$ the cluster expansion yields the excess number of electrons per a single atom (not equal to the nuclear charge). Specifically, in the limit of weak non-ideality and weak electron degeneracy considered one gets $N_{e}=Z+Z_{0}/(Z_{0}+1)$: that is, in the limit of complete ionization ($Z_{0}=Z$) $N_{e}$ is overestimated by a factor of $(1+1/(Z+1))$. It may be shown that in the complete-ionization limit Eq. (\ref{cluster_energy}) overestimates electronic pressure and energy in just the same way.

Here, one more fact should be noted. The derivation of Eq. (\ref{chem_cond}) did not assume that the function $g(r)$ is also dependent on the free-electron density $n_{e}^{0}$. If this dependence is taken into account, one can show that Eq. (\ref{chem_cond}) should be written with the substitution:
\begin{equation}\label{g_subst}
g(r)\to g(r)+n_{e}^{0}\,\dfrac{\partial g(r)}{\partial n_{e}^{0}}
\end{equation}
Substituting the asymptotics of the pair correlation function (\ref{h_res}) and its derivative
\begin{equation}\label{g_deriv}
n_{e}^{0}\,\dfrac{\partial g(r)}{\partial n_{e}^{0}}=\beta\,Z_{0}^{2}\,e^{-r/r_{D}}\left(-\dfrac{2}{r}+\dfrac{Z_{0}+\frac{1}{2}}{(Z_{0}+1)\,r_{D}}\right)
\end{equation}
into the equation following from Eq. (\ref{chem_cond}) with the substitution (\ref{g_subst}) performed, we find that the left-hand side of the equation remains unchanged (with small corrections omitted) while the right-hand side takes the form:
\begin{equation}\label{chem_cond_rhs_mod}
\dfrac{\mu_{e}}{\left(Z_{0}+1\right)^{2}n_{i}}.
\end{equation}
Though Eq. (\ref{chem_cond_rhs_mod}) as well as Eq. (\ref{chem_cond_rhs}) are not exactly coincident with Eq. (\ref{chem_cond_lhs}), this result shows that allowing for the dependence of the pair correlation function on the free-electron density $n_{e}^{0}$ may substantially affect the variational condition for the chemical potential.

Finally, we note one more result immediately following from the considerations above. As the electron transport coefficients may be expressed through the transport electron-scattering cross-section $\sigma_{tr}(\eps)$, in the Born approximation, valid at high free electron energies ($\eps\gg Z_{0}^{2}$) and/or high densities $\left(\rho\left(\text{g/cm}^{3}\right)\gg A\,Z_{0}^{3}\right)$, one gets \cite{Starrett2017_2}
\begin{equation}\label{cross_born}
\sigma_{tr}(\eps)=\dfrac{1}{4\pi^{2}}\int\limits\left(V^{PA}(q)\right)^{2}S(q)\,(1-\cos\theta)\,d\Omega,
\end{equation}
where $q=2\sqrt{\eps\,(1-\cos\theta)}$, $S(q)=1+n_{i}\,h(q)$ is the ion structure factor. According to Eq. (\ref{h_res}), in the limit of weak non-ideality and weak electron degeneracy we find that \begin{equation}\label{structure}
S(q)=\dfrac{q^{2}+1/d^{2}}{q^{2}+1/r_{D}^{2}}.
\end{equation}
Substituting the ion structure factor (\ref{structure}) and the Fourier transform of the pseudoatom potential (\ref{screening_res}) into Eq. (\ref{cross_born}) we get
\begin{equation}\label{transport}
\sigma_{tr}(\eps)=\pi\,Z_{0}^{2}\,\Lambda(\eps)/\eps^{2}
\end{equation}
with the Coulomb logarithm
\begin{equation}\label{Coulomb}
\Lambda(\eps)=\frac{1}{2}\left(1+\dfrac{1}{Z_{0}}\right)\ln\left(1+\left(\dfrac{2r_{D}}{\lambda}\right)^{2}\right)-
\dfrac{1}{2Z_{0}}\,\ln\left(1+\left(\dfrac{2d}{\lambda}\right)^{2}\right),
\end{equation}
where $\lambda=1/\sqrt{2\eps}$.

It is proposed in Ref. \cite{Starrett2017_2} that the electron scattering cross-section averaged over various ion positions may be substituted for the scattering cross-section in the average potential due to all plasma particles as calculated with the Starrett and Saumon model. If the Born approximation is valid, this approach results in the cross-section (\ref{transport}) with a Coulomb logarithm
\begin{equation}\label{Coulomb_rD}
\Lambda(\eps)=\dfrac{1}{2}\left(\ln\left(1+\left(\dfrac{2r_{D}}{\lambda}\right)^{2}\right)-\dfrac{4r_{D}^{2}}{4r_{D}^{2}+\lambda^{2}}\right).
\end{equation}
Though Eq. (\ref{Coulomb_rD}) differs from the more accurate Eq. (\ref{Coulomb}), one can readily show that under the condition $r_{D}\gg\lambda$ these expressions coincide with logarithmic accuracy both in the $Z_{0}\gg1$ and $Z_{0}\sim1$ cases. Therefore, the method to calculate transport coefficients proposed in Ref. \cite{Starrett2017_2} provides (with logarithmic accuracy) correct asymptotics in the limit of weak non-ideality and weak electron degeneracy.




\bibliographystyle{UNSRTRM}
\bibliography{references}


\end{document}